\documentstyle[11pt,newpasp,twoside,epsf]{article}
\def\edcomment#1{\iffalse\marginpar{\raggedright\sl#1\/}\else\relax\fi}
\reversemarginpar

\begin{document}
\title{Properties of Black Holes in Stellar Binary Systems}
\author{R. A. Remillard}
\affil{Massachusetts Institute of Technology, 
Room 37-595, Cambridge MA 02139, USA}

\begin{abstract}
Recent X-ray timing and spectral observations of black hole binaries
in outburst have redefined methods for investigating the properties of
Galactic black holes and the physics of accretion flows. High-frequency
X-ray QPOs in 5 systems (60--300 Hz) continue to be investigated as a
possible means to constrain black hole mass and spin via
interpretation as oscillations due to strong-field effects in General
Relativity.  In principle, black hole mass and spin may also be
constrained via determination of the inner radius of the accretion
disk, using spectroscopic parallax of the X-ray thermal component.
However, an accurate application of this technique is fraught with
difficulties. Monitoring programs that track the evolution of the
thermal and power-law components in the X-ray spectrum provide new
insights into modes of energy flow and the formation of relativistic
jets.
\end{abstract}

\section{Introduction}

During the 1980's and 1990's, optical radial velocity studies of
stellar companions in quiescent X-ray binary systems in the Galaxy
identified a group of compact objects that were more massive than the
theoretical upper limit ($\sim3$ M\sun) for the mass of a neutron star
(see Charles 1998).  It was noted that while in outburst, these
``heavy'' compact objects also lack the X-ray signatures of neutron
stars, viz. pulsations or type I X-ray bursts.  These cases eventually
gained widespread recognition as ``dynamically established black hole
binaries''.  The measured optical mass functions were combined with
estimates for the binary mass ratio and inclination angle, which were
also determined from optical studies of the companion stars, to yield
black hole mass estimates in the range of 5--15 M\sun.  Recent
additions to this group include SAX~J1819.3$-$2525 (Orosz et
al. 2001), and XTE~J1118+480 (McClintock et al. 2000). There are now
13 known X-ray binaries in which the compact object is very likely to
be a black hole. During X-ray outbursts, the black-hole systems often
exhibit composite spectra consisting of a $\sim 1$ keV thermal
component and a broad power-law component that may extend well beyond
100 keV (see Tanaka \& Lewin 1995). These spectral characteristics are
often used to suggest that a particular source is a black hole
``candidate'' (BHC) when dynamical observations of the companion star
are lacking, sometimes because of high extinction at optical/IR
wavelengths.

Another important topic for quiescent X-ray binaries is whether the
black hole systems are further distinguished (compared to neutron
stars) by spectral characteristics that can only be understood as a
consequence of a black hole's event horizon (Narayan, Garcia, \&
McClintock 1997; Menou et al. 1999).  The optical, UV, and X-ray
energy distribution of quiescent black hole binaries is inconsistent
with models for thermal emission from a disk with a low rate of mass
transfer.  The high-energy radiation also appears intrinsically
fainter than that of quiescent neutron-stars (McClintock \& Remillard
2000). The differences between black holes and neutron stars may then
be explained as the inevitable thermalization of infall energy
(i.e. crash) at the neutron star surface, in contrast with the flow of
energy into a black hole's event horizon.  While the qualitative
aspects of quiescent spectra in black hole binaries is quite
intriguing, the application of detailed models remains
controversial. The ``Advection Dominated Accretion Flow'' model
(e.g. Narayan \& Yi 1995) can account for many of the observed
spectra, but issues of convective stability have ushered in the
``Convection Dominated Accretion Flow'' model (Ball, Narayan, \&
Quataert 2001). There are also alternative scenarios for explaining
the ``underluminous'' spectra of quiescent disks, e.g. by via mass
loss from the system (Blandford \& Begelman 1999) or by invoking
magnetic instabilities in a cool disk (Nayakshin 2000).

One undesirable consequence of low radiative efficiency in quiescent
transients is the limited opportunity to study black holes by
observing the detailed characteristics of the accreting matter {\it
near} the event horizon.  This perspective has encouraged renewed
interest in X-ray outbursts, where the highest accretion rates are
seen. Since the the launch of the {\it Rossi X-ray Timing Explorer}
(RXTE) in 1995 December, we have witnessed historical contributions to
scientific studies of black holes in active states of accretion. These
efforts have brought new themes to the forefront of high-energy
astrophysics, such as the application of General Relativity (GR) to
black hole binaries.  In the sections below, we review some of the
ongoing efforts to use X-ray timing and spectroscopy to evaluate black
hole mass and spin, to understand the physics of black hole accretion,
and to determine the origins of relativistic jets in X-ray binary
systems (i.e. the ``microquasars'').

\subsection{X-ray Spectra of Accreting Black Holes}

The dedication of the RXTE Mission to studies of
X-ray transients is reflected in the statistics (1996-2000) of
monitoring programs for pointed observations: 81 exposures of
GRO~J1655$-$40 (1996-1998); 222 for 4U1630$-$47 (4 outbursts), 316 for
XTE J1550$-$564 (3 maxima), and 618 for GRS 1915+105. The typical
observation with the PCA (2-60 keV) and HEXTE (20-200 keV) instruments
provides several million detections during a few ks with sub-ms time
resolution. The All Sky Monitor (ASM) independently records X-ray
intensities (2-12 keV) by scanning the celestial sphere several times
per day.

\begin{figure}
\plotone{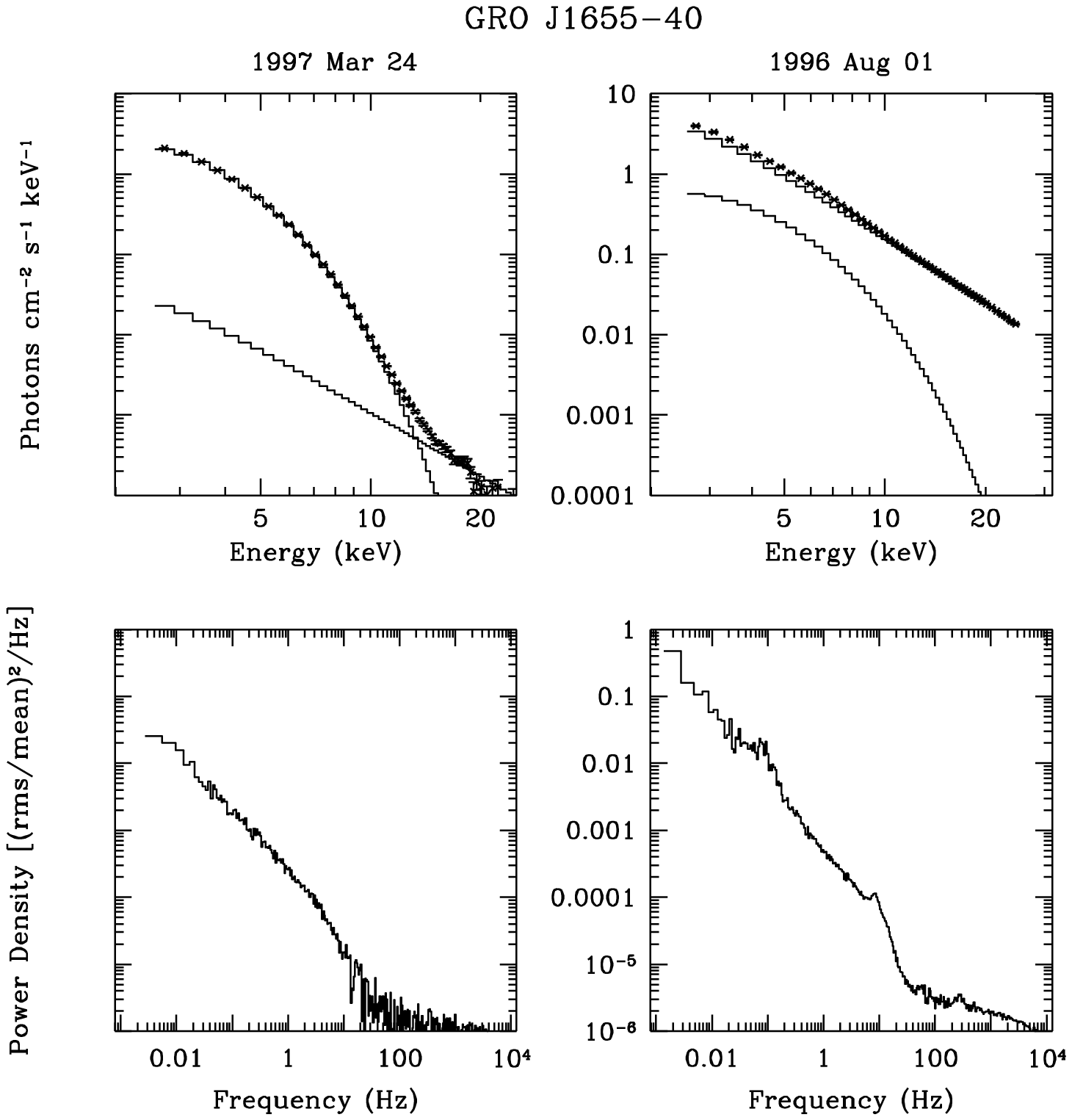}
\caption{X-ray spectrum and power density spectrum of GRO J1655$-$40
in the ``high state'' (1997 Mar 24) and the ``very high state'' 
(1996 Aug 1). The very high state corresponds to episodes of highest 
luminosity, where the spectrum is dominated by a non-thermal X-ray 
power-law. In this source, QPOs at 0.1, 8, and 300 Hz are visible 
in the very high state. }
\end{figure}

Black hole binaries in states of active accretion usually show two
spectral components with a large diversity in their relative strength.
This is shown in Figure 1 for the case of GRO J1655$-$40 The ``soft''
X-ray component represents thermal emission ($\sim 1$ keV) from an
accretion disk, while the power-law continuum in hard X-rays is
commonly associated with inverse Compton scattering of ambient photons
by energetic electrons. As the luminosity of GRO J1655$-$40 exceeds
$\sim$0.2 $L_{Edd}$, the power-law becomes dominant and several types
of quasi-periodic oscillations (QPO) are seen in the PCA power
spectrum (see lower panels). The nature of this Compton emission,
which is the dominant mechanism for radiative losses at the highest
accretion rates, remains one of the fundamental mysteries of accretion
energetics in black hole binaries.

At luminosities of roughly $\sim 0.01-0.2 L_{Edd}$, the thermal
component (accretion disk) usually dominates the spectrum.  Such
``soft state'' or ``high state'' spectra have been modelled with a
multi-temperature accretion disk (Shakura \& Sunyaev 1973; Mitsuda et
al. 1984) with an inner radius near the marginally stable orbit
predicted by GR theory. Here, a weaker Compton component is usually
present, but its origin and geometry are unknown. As the luminosity
evolves, unexpected variations are seen in the apparent disk
temperature and radius for many sources (e.g. Sobczak et
al. 1999a). The interpretation of these results requires new disk
models with more sophisticated treatment of effects due to GR and
radiative transfer through the disk atmosphere (see 2.2 below).

We continue to learn about the importance of the hard X-ray power law,
but its behavior is fraught with complications.  When the spectral
characteristics and evolution of all black hole binaries is
considered, it is found that the power-law component has two
inherently different types (Grove et al. 1998). The spectrum
associated with the very high state (Fig. 1) has a photon index
$\sim$2.5, and this component has been seen out to 1 MeV or higher
without a cutoff in a few bright sources (Tomsick et al. 1999).  A
different power-law at lower luminosity (roughly $L_x < 0.01
L_{Edd}$), with a photon index $\sim$1.7 and an exponential cutoff
near 100 keV, characterizes the ``low-hard'' state (e.g. Cyg X$-$1).
Since the high-energy cutoff suggests a limit to the energy
distribution of Compton electrons, it seems paradoxical that the
low-hard state is associated with a steady type of radio jet (see
below).

There are now considerable efforts underway (e.g. Hua, Kazanas \& Cui
1999; Poutanen \& Fabian 1999; Nowak et al. 1999).  to deduce physical
insights about the hard X-ray component by combining spectral analyses
with variability measurements such as phase lags and coherence
(vs. Fourier frequency) between different energy bands, as well as the
properties of low-frequency QPOs. These oscillations can attain
amplitudes as high as 30-40\% of the average flux, e.g. at 1-5 Hz
(Morgan, Remillard, \& Greiner 1996; Cui et al. 1999).

Further insights can be gained by identifying repetitive patterns of
correlated spectral and timing variations gained from the monitoring
programs. Prior to RXTE is had been widely presumed that the emergent
photon spectrum depends primarily on the black hole mass and the
accretion rate.  Then it was shown that the hard/soft state transition
in Cyg X$-$1 involves almost no change in bolometric luminosity (Zhang
et al. 1997). In a more general sense, it has been noted that almost
all of the BHCs with persistent X-ray emission exhibit the same
spectral states as X-ray transients, while extreme changes in mass
accretion rate are only expected for the latter group (Wilms et
al. 2001). Along the same lines, it has been shown that changes in
luminosity in a single source, in different outbursts, may follow very
different tracks in the way the energy divides between the accretion
disk and power-law components.  In the microquasar GRS~1915+105, a
plot of the luminosity contributions from the disk vs. power-law
components reveals two tracks: one branch shows luminosity changes
occurring primarily in the disk, while an alternate track shows
luminosity changes entirely diverted to the power-law.  The latter
track contains {\it all} of the observations in which there are
low-frequency (0.5 to 15 Hz) QPOs (Muno, Morgan, \& Remillard 1999).
Even more complicated tracks are seen in the BHC XTE~J1550$-$564
(Sobczak et al. 1999b). All of these results imply that there are
primary factors that have received insufficient recognition in our
descriptions of the energetics of black hole accretion. The dependent
variables may include conditions in the outer disk,
magneto-hydrodynamic turbulence, or the presence of jets.

\section{Constraints on Black Hole Properties Using General Relativity}

\subsection{High-Frequency Quasi-Periodic Oscillations in X-rays}

The X-ray sources GRS~1915+105 and GRO~J1655$-$40 yielded the first
detections of fast X-ray oscillations thought to be signatures of
strong-field GR effects in the inner accretion disk. RXTE discovered a
transient QPO in GRS~1915+105 at 67 Hz (Morgan, Remillard, \& Greiner
1997). When detected, the QPO frequency appears fixed, despite
variations in X-ray luminosity by a factor of 4. A QPO at 300 Hz in
GRO~J1655$-$40 (Remillard et al. 1999b) was seen during $\sim$6
observations that exhibit the highest levels of luminosity observed by
RXTE from this source. The first evidence of a BHC with a fast QPO
having variable frequency is found in the case of XTE~J1550$-$564
(100-284 Hz; Remillard et al. 1999a; Homan et al. 2001). Additional
high-frequency QPOs were found in the recurrent BHC transient
4U~1630$-$47 (184 Hz; Remillard et al. 1999c), and in XTE~J1859+226
(150-187 Hz; Cui et al. 2000), again by combining power spectra over
many PCA observations at high luminosity.  Figure 2 illustrates these
results.  All of these QPOs have rms amplitudes in the range of 0.5\%
to 3.5\% of the mean count rate.

\begin{figure}
\plotone{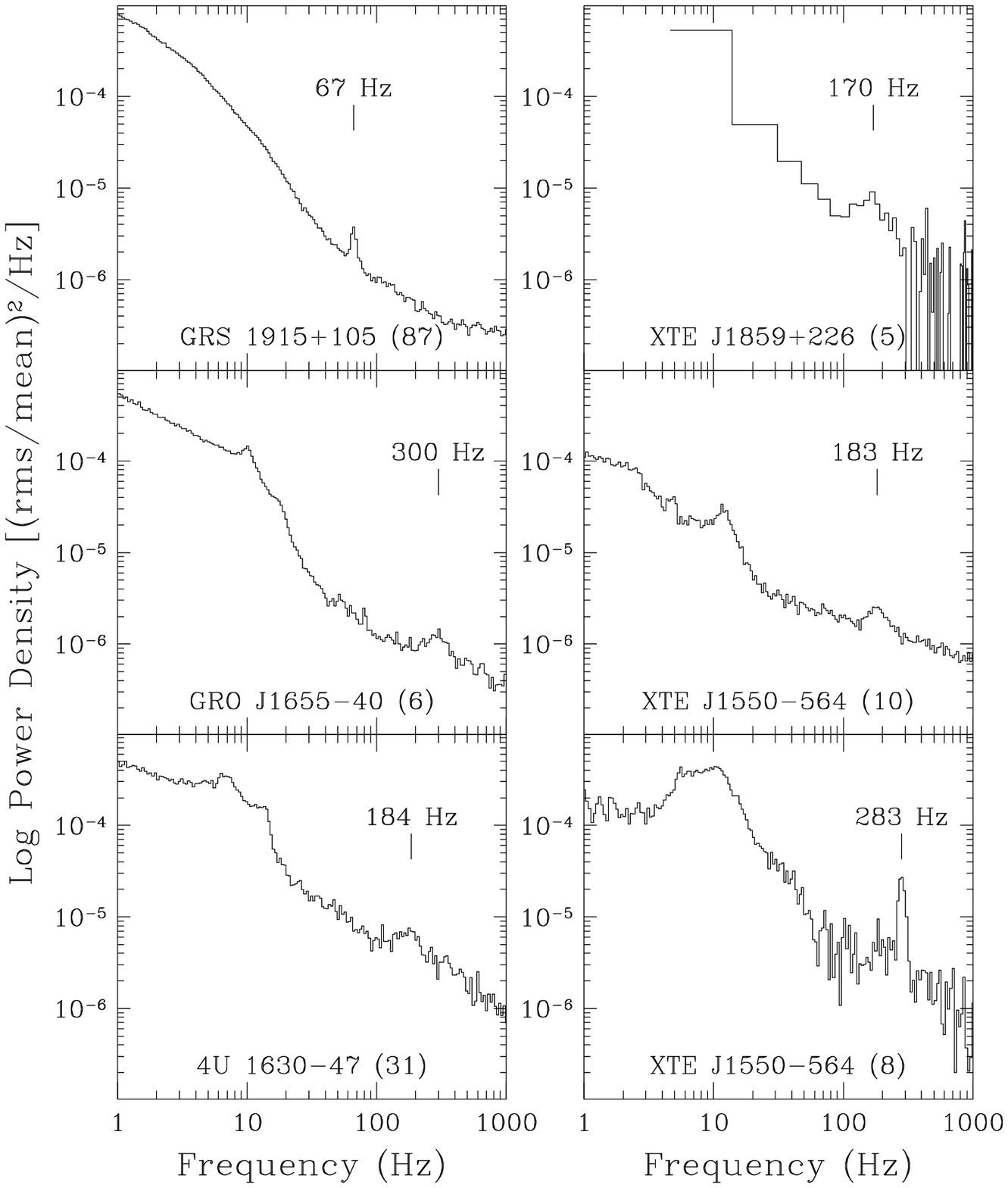}
\caption{High frequency QPOs in 5 black hole candidates.
Two observations for XTE J1550$-$564 are shown to illustrate the large
variations in frequency and coherence for the fast QPO from this
source. On the opposite extreme, the 67 Hz QPO in GRS 1915+105
is confined to the range of $\pm 2$ Hz during $\sim$30 observations
when a fast QPO was detected.}
\end{figure}

Given those high frequencies and their display in X-rays, it is
straightforward to hypothesize that the fast QPOs are produced in the
inner accretion disk, near the black hole event horizon. Here, there
are GR predictions of oscillations that do not exist in Newtonian
gravity.  The competing models for these oscillations incorporate GR
effects at some level, and the QPO frequency then depends on both the
mass and spin of the black hole, and possibly on the gas conditions or
vertical structure of the inner disk.  Some models invoke GR phenomena
directly, suggesting that the QPOs represent the the last stable orbit
(Shapiro \& Teukolsky 1983), or ``frame dragging''
(i.e. Lense-Thirring precession; Merloni et al. 1999), or relativistic
periastron precession (Stella, Vietri, \& Morsink 1999). Other models
suggest a more subtle interplay between GR and fluid dynamics, such as
''disko-seismic modes'' (Perez et al. 1997; Ortega-Rodriguez \&
Wagoner 2000) or inertial-acoustic instabilities (Milsom \& Taam
1997). To date, there is only one of the five sources (GRO~J1655$-$40)
for which the mass of the black hole ($\sim 7$ M\sun ) is available
from optical studies (Orosz \& Bailyn 1997; Shahbaz et al. 1999). In
this case the dimensionless spin parameter ($a$) could be as low as
$a=0$ if the QPO signifies the last stable orbit ($R \sim 6 GM/c^2$),
while values closer to $a=1$ are deduced for several of the other
models. 

The possibility that one may constrain the mass and spin of an
accreting black hole using high-frequency X-ray QPOs, which are
independent of the usual systematic uncertainties about distance,
inclination angle, and extinction, is surely a compelling theme for
vigorous scientific analyses. However, there are many serious problems
that must be addressed. All of the high-frequency QPOs appear to be
stronger at higher photon energy, which resembles the X-ray power law,
rather than the thermal component from the inner disk, where the QPOs
are expected to originate. Therefore, detailed emissivity models are
required for each type of GR oscillation, including considerations for
the energetic electrons believed responsible for the X-ray power-law
via inverse Compton emission. We have seen the first simulations of
frame dragging oscillations and how they may survive against damping
(Markovic \& Lamb 1998). There are also more generalized
considerations as to how these oscillations may be excited by the
hydrodynamic turbulence in the accretion flow (Psaltis \& Norman
2001).  All of these crucial topics will continue to evolve in
parallel with observational developments gained with RXTE.

\subsection{Spectroscopic Inference of the Inner Radius of the Disk}

Spectral analysis of the thermal component in the X-ray spectrum
yields information on the radius and temperature of the inner disk,
when the distance and binary inclination are well constrained, e.g. by
optical and radio observations. If the accretion disk extends all the
way in toward the event horizon, i.e. there are no shocks or
transitions to radial accretion, then the disk is expected to
terminate at the radius of the last stable orbit given by GR
theory. Since the luminosity is dominated by the inner disk, there is
an opportunity, in principle, to derive an absolute value for the
inner disk radius and then to constrain the value of the black hole's
spin if the mass is known via binary dynamics. This is tantamount to a
''spectroscopic parallax'' of the inner accretion disk.
 
Zhang, Cui, \& Chen (1997) applied this technique to observations of
microquasars, while estimating corrections for effects such as
radiative transfer through the disk atmosphere and GR modifications on
the structure and emissivity of the inner disk.  The accuracy of these
corrections was strongly questioned (Merloni, Fabian, \& Ross
2000). There are also complications arising from unexplained changes
in the apparent radius of the inner disk in several sources (Belloni
et al. 1997; Sobczak et al. 1999b; Muno, Morgan, \& Remillard
1999). Finally, a number of black hole accretion models specify
structural elements that contradict the assumption of an idealized
accretion disk that extends all the way to the event
horizon (e.g. Chakrabarti 1996; Titarchuk, Lapidus, \& Muslimov 1998).

Despite these problems, there may eventually be useful ways to derive
constraints on black hole mass and spin from X-ray spectroscopy. 
Progress will depend, in part, on improved accretion disk models. 
One strategy might be to select only the spectra that exhibit a 
dominance of the thermal component, with perhaps an 
additional selection for minimal variability displayed in the power
spectrum.  It is also conceivable that accretion models could gain
important diagnostic support from X-ray line spectroscopy, if
observations from the Chandra Observatory or an anticipated re-flight
of the Astro-E calorimeter establish broad Fe lines in the spectra of
black hole binaries.

\section{ The Disk-Jet Connection \label{subsubsec:jetbh}}

Since 1994, radio astronomers have discovered a handful of Galactic
X-ray binaries that produce transient radio jets traveling at
velocities $ \sim 0.9 $c (Mirabel \& Rodgriques 1999).  These sources are
referred to as ``microquasars'', since they appear to be scaled-down
versions of the jets in AGN.  Interest in microquasars is heightened
by the opportunity to observe the formation and evolution of mass
ejections on timescales of seconds to days, while such processes take
years to millennia in quasars. Theoretical arguments favor magnetic
fields as the mechanism that collimates and accelerates the
ejecta. The outflow energy may originate in the spin of the black hole
and/or the accretion energy in the disk (see Blandford 2000).  In some
microquasars, one may study the jets with knowledge of the black hole
mass, which is constrained via optical studies of the mass-donor
companion star.

One may describe microquasar jets as occurring in at least 4 types:
explosive bipolar ejections at relativistic speed; weaker oscillatory
ejections at regular intervals ($\sim 30$ min); very weak radio flares
that preceed X-ray flares, and steady jets which may persist from days
to months. It can be argued that GRS~1915+105 and Cyg X$-$3 are the
prototypical microquasars for their recurrent production of jets of
more than one type. The X-ray (ASM) and radio (Greenbank
Interferometer) light curves for these sources are shown in Figure 3.

\begin{figure}
\plotone{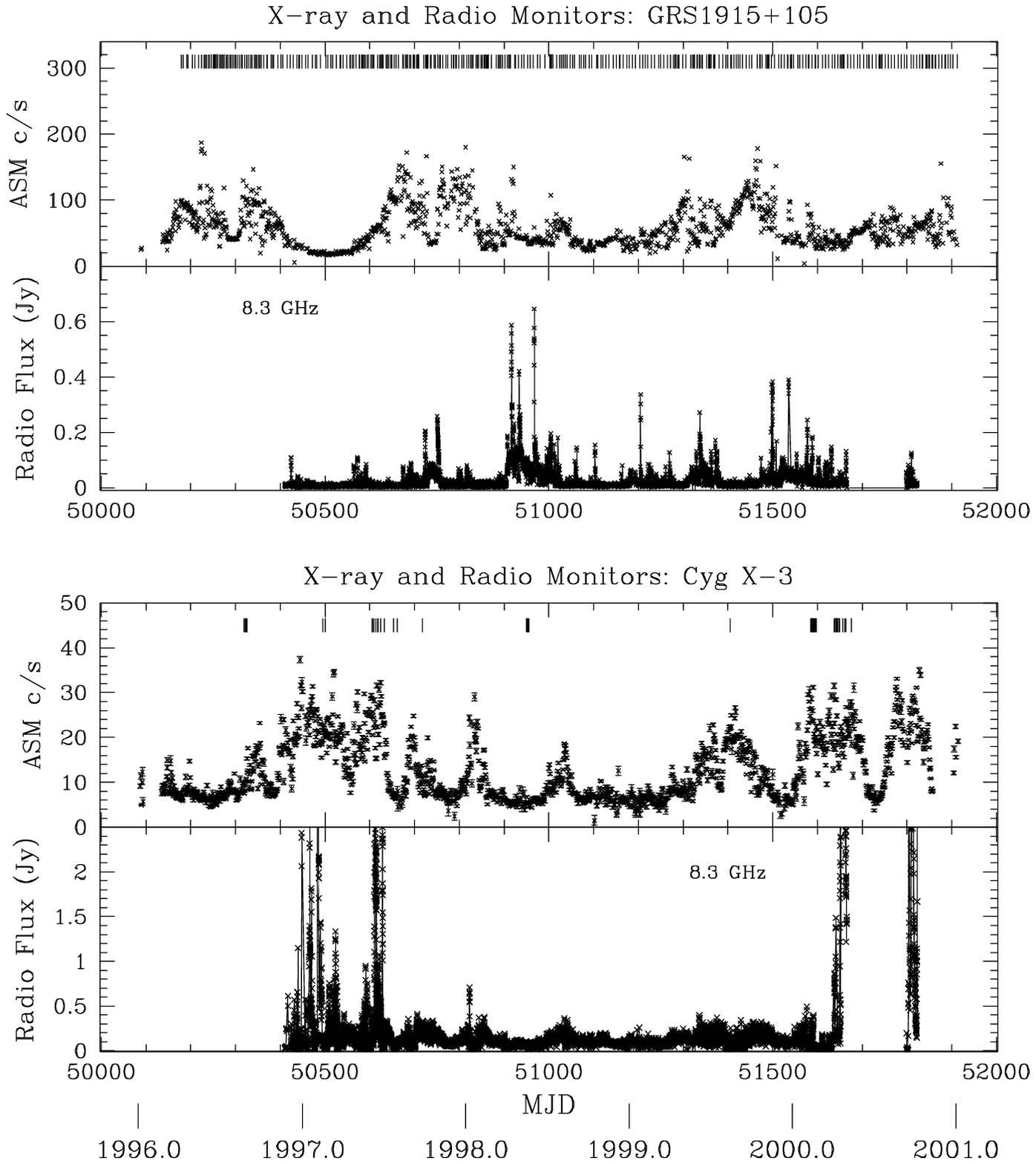}
\caption{X-ray and radio monitoring observations of the most active 
microquasars: GRS~1915+105 and Cyg X$-$3. The X-ray light curves
are binned in 1 day intervals, and the rows vertical ticks above
them show the times of RXTE pointed observations.}
\end{figure}

In the case of GRS~1915+105, the ''superluminal'' ejections seen with
the VLA and MERLIN (Mirabel \& Rodriguez 1994; Fender et al. 1999a)
typically reach maxima of 0.5--2.0 Jy at 2.25 GHz. Even stronger
events are seen in Cyg X$-$3 (e.g. Newell et al, 1998). And in the
case of V4641 Sgr (Hjellming et al. 2000), a dynamical black hole
binary with an evolved A star companion (Orosz et al. 2001), the
fast-moving, single-sided jet implies ejection at
0.99 c. The X-ray behavior associated with these rare
events is not known. However, weaker ejections (0.01--0.10 Jy) may
appear as series of radio flares that repeat every $\sim 30$ min
(Fender \& Pooley 1998; Eikenberry et al. 1998; Fender \& Pooley
2000). These oscillatory ejections have brought important lessons
about the crucial role of X-ray monitors in understanding the jet
formation process.  There are smooth IR and radio flares (to 0.02 Jy
at 2.25 GHz) that represent successively delayed peaks in synchrotron
emission associated with the expansion of an ejected plasma cloud. The
X-ray light curve is far more complex, and yet the wild patterns also
repeat every 30 min!  Eikenberry et al. (1998) found similar sequences
of X-ray dips, flashes, and bright flares preceeding all 6 IR flares
observed at both wavelengths. Spectral analysis of the X-ray dip
indicates evolution from a small, hot disk to a much cooler one with
larger inner radius (Belloni et al. 1997). The conclusion is that the
emergence of the jet is a consequence of the rapid disappearance and
followup replenishment of the inner accretion disk.

There are also reports of weaker, isolated radio or IR flares that
preceed X-ray flares (Eikenberry et al. 1999; Feroci et al. 1999).
These are interpreted as small ejections that arise from
''outside-in'' instabilities, or waves that fuel a small jet before
reaching the innermost region of the accretion disk.

Finally, there has been substantial recent progress in demonstrating
the presence of a steady type of jet associated with the ''low-hard''
X-ray state seen in many X-ray binaries.
Radio monitoring and ASM observations of GX 339$-$4 have revealed
positive correlations during the X-ray hard state, with the radio
flux becoming quenched as GX 339$-$4 transits to its soft state
(Hannikainen et al. 1998; Corbel et al. 2000). The radio emission is
consistent with optically-thick synchrotron emission of extent $>
10^{12}$\,cm (Wilms et al. 1999).  A clear association of the hard
state with a steady type of jet was gained by Fender et al. (1999b),
with the detection of extended radio structure during the hard state.
Simultaneous RXTE observations indicate an X-ray power law ($Gamma
\sim 1.7$) that is consistent with Comptonization via a 50-100 keV
corona with radius approximately $< 100~GM/c^2$. 

There are also VLBI images of a jet associated with steady X-ray and
radio emission in the case of GRS~1915+105 (Dhawan, Mirabel, \&
Rodriguez 2000). Here the jet is visible on all dimensions between 10
and 500 AU! In addition, steady-jet properties are seen
in the low-hard state of Cyg X$-$1 (Pooley, Fender, \& Brocksopp 1999;
Brocksopp et al. 1999). All of these results are focusing efforts to
relate the ($Gamma \sim 1.7$) type of power law to Compton scattering
at the base of a steady jet. As might be expected, however, the
detailed changes in X-ray characteristics as a function of radio flux
within the low-hard state appear to be quite complicated (Muno,
Remillard, \& Morgan 2001).

In closing, it must be noted that the classical work to determine
binary parameters and stellar constituents for the Galactic black hole
systems remains a core component of this science.  Confronting the
high-energy phenomenology of these systems requires the knowledge and
context provided by measurements of the binary system, and any
practical prospects for gaining evaluations of black hole spin require
the mass measurements derived from binary dynamics. As we continue to
pursue all of the science of black hole mass and spin and the jets
that propel matter at relativistic speeds directly from the
gravitational jaws of these black holes, it is essential that we
maintain a vigorous program of ground-based observations for the same
X-ray transients that are studied from space.

\end{document}